\documentclass[superscriptaddress,twocolumn,showpacs,amsmath,amssymb]{revtex4}

\usepackage[dvips]{graphicx}
\usepackage{graphicx}

\begin{document}

\title{
Quasi-elastic scattering in the $^{20}$Ne + $^{90,92}$Zr reactions:
role of non-collective excitations
}

\author{S. Yusa}
\affiliation{
Department of Physics, Tohoku University, Sendai 980-8578,  Japan} 

\author{K. Hagino}
\affiliation{
Department of Physics, Tohoku University, Sendai 980-8578,  Japan} 

\author{N. Rowley}
\affiliation{
Institut de Physique Nucl\'{e}aire, UMR 8608, CNRS-IN2P3 et Universit\'{e}
de Paris Sud, 91406 Orsay Cedex, France}

\begin{abstract}
Conventional coupled-channels analyses, that take account of only the
collective excitations of the colliding nuclei, have failed to reproduce
the different behavior of the experimental 
quasi-elastic barrier distributions for the $^{20}$Ne + $^{90,92}$Zr systems. 
To clarify the origins of this difference,
we investigate the effect of non-collective excitations of the Zr isotopes. 
Describing these excitations in a random-matrix model,
we explicitly take them into account 
in our coupled-channels calculations.
The non-collective excitations are capable of reproducing
the observed smearing of the peak structure in the barrier distribution
for $^{20}$Ne + $^{92}$Zr, while not significantly altering the structure 
observed in the $^{20}$Ne + $^{90}$Zr system. The difference is essentially 
related to the closed neutron shell in $^{90}$Zr.
\end{abstract}


\maketitle

\section{Introduction}

In heavy-ion reactions around the Coulomb barrier,
the relative motion 
between the projectile and the target nuclei 
couples to the internal degrees of freedom of the 
colliding nuclei in a decisive way, leading to a distribution of
potential-barrier heights around the energy of the uncoupled Coulomb
barrier~\cite{DLW83}.
This leads to the strong enhancements of subbarrier fusion cross sections
observed in a number of medium-heavy systems~\cite{dasgupta,BT98,HT12},
when compared with the predictions of a simple potential model.
The distribution of potential barriers $D_{\rm fus}$ can be obtained from measured
fusion cross sections $\sigma_{\rm fus}(E)$ by taking the second derivative
$D_{\rm fus} = d^2\left({E\sigma_{\rm fus}}\right)/dE^2$
with respect to the incident energy~$E$~\cite{dasgupta,RSS91,L95}.
The resulting fusion barrier distributions often exhibit 
structures~\cite{dasgupta,L95,LRL93} that are characteristic of the
details of the collective states to which the entrance channel can couple.

It has been recognized that the concept of a barrier distribution
can also be applied to  cross sections for heavy-ion, quasi-elastic scattering
(that is, to the sum
of elastic and inelastic scattering and transfer cross sections)
\cite{timmers,HR04}. 
For these processes, one can extract the 
barrier distribution from the measured total
differential cross section $\sigma_{\rm qel}(\theta)$  
at backward angles using the simple formula 
$D_{\rm qel} = -d\left(\sigma_{\rm qel}
/\sigma_{\rm R}\right)/dE$. Notionally, $\sigma_{\rm qel}$ 
and the Rutherford cross section $\sigma_{\rm R}$ should be taken 
at a scattering angle $\theta=\pi$, though any large angle may be used
with an appropriately defined `effective' energy (see Sec.~III).
This quasi-elastic barrier distribution is also sensitive to coupling
effects, and behaves in a very similar way to 
that for fusion~\cite{timmers,HR04,zamrun}.
Note that $\sigma_{\rm fus}$ and $\sigma_{\rm qel}$ are
in some sense complementary to one another, 
in that fusion corresponds to
penetration of the potential barrier, whereas back
scattering corresponds to reflection from the barrier.

In order to take account of coupling effects in the reaction process, 
the coupled-channels method is considered to be a
standard approach~\cite{HT12,ccfull}. 
Conventionally, only a few low-lying collective excitations,
such as surface vibrations of spherical nuclei or 
rotations of nuclei with static deformations, have been taken into account. 
These coupled-channels analyses have successfully accounted for 
the strong enhancement of subbarrier fusion cross sections 
as well as for the observed structures in the barrier distributions
for many systems~\cite{HT12}.

Nevertheless, there remain several challenging problems 
to be explored in the present coupled-channels approach. 
For instance, it has been a long-standing problem that 
a standard value of $a\sim 0.63$~fm for the surface diffuseness 
parameter of the real nuclear potential appears too small to account for 
fusion data, even though this value is required to fit scattering data \cite{MHD07,N04}.
This problem is also related~\cite{HRD03,ichikawa,esbensen}
to the deviations of fusion cross sections at deep subbarrier energies 
from the predictions of standard coupled-channels calculations~\cite{J02,JRJ04,D07,S08}.

Another example is quasi-elastic scattering in the $^{20}$Ne +
$^{90,92}$Zr systems~\cite{piasecki}. 
For these systems, 
the experimental quasi-elastic barrier distributions
exhibit significantly different behavior, that is,
the barrier distribution for the $^{20}$Ne + $^{92}$Zr system shows a 
much more smeared structure than that for the $^{20}$Ne + $^{90}$Zr system. 
In contrast, the coupled-channels calculations that include the collective excitations 
in the $^{20}$Ne and Zr isotopes, yield very similar barrier distributions 
for the two systems. 
In the calculations, the rotational excitations of the 
strongly deformed $^{20}$Ne nucleus 
dominate the barrier structure, and the collective vibrational 
excitations in the Zr isotopes are found to play a minor role.
Experimental data for the 
total transfer cross section at an energy near the Coulomb barrier
have been found to be essentially the same~\cite{piasecki},
and the difference in the barrier distributions has been 
conjectured to originate
from differences in the non-collective excitations in the two Zr isotopes.
In fact, since the $^{92}$Zr nucleus has two neutrons outside 
the $N = 50$ closed shell in $^{90}$Zr, a larger number
of non-collective excited states are present in the spectrum 
(for example, the number of known states up to 5~MeV 
is only 35 for $^{90}$Zr but 87 for $^{92}$Zr~\cite{bnl}). 

The aim of this paper is to investigate whether the non-collective 
excitations of the $^{90,92}$Zr isotopes can explain the observed differences 
in the quasi-elastic barrier distributions for the 
$^{20}$Ne + $^{90,92}$Zr systems, 
by explicitly taking them into account 
in large-scale coupled-channels calculations.
In order to describe the non-collective 
excitations, we employ the random-matrix model, which 
was originally introduced in the 1970's by Weidenm\"uller {\it et al.} 
in order to study deep-inelastic collisions 
\cite{KPW76,AKW78,BSW78,akw1,akw2,akw3,akw4}.
In Ref.~\cite{YHR13},  
we have shown that the non-collective excitations are not sensitive to 
how they are modeled and that the random-matrix method provides 
a good way to treat them when the relevant properties of the non-collective states 
are not well known (see also Refs.~\cite{YHR10,YHR12}). 
This justifies the use of the random-matrix model in the present analyses. 

The paper is organized as follows. 
In Sec.~II, we present the coupled-channels formalism and its various ingredients.
In particular we detail the collective coupling parameters and the generation of
the random-matrix, non-collective couplings that will be applied to the 
quasi-elastic scattering in the $^{20}$Ne + $^{90,92}$Zr systems.
In Sec.~III, we discuss the effect of the non-collective excitations on the 
corresponding quasi-elastic scattering cross sections, on the barrier
distributions, and on the $Q$-value distributions.
The paper is summarised in Sec.~IV.

\section{Coupled-channels method with non-collective excitation}

\subsection{Coupled-channels equations}

The coupled-channels equations in the isocentrifugal approximation
are given by~\cite{HT12},
\begin{eqnarray}
 \left[-\frac{\hbar^2}{2\mu}\frac{d^2}{dr^2} +  \frac{J(J+1)\hbar^2}{2\mu r^2}
+ V_{\rm rel}(r) + \epsilon_n - E\right]u_n^{J}(r) \nonumber&&\\
 +  \sum_m V_{nm}(r)u_m^{J}(r) = 0,
\label{cceq}
\end{eqnarray}
where $\epsilon_n$ is the excitation energy for the $n$-th channel
and $J$ is the total angular momentum. 
$\mu$ and $V_{\rm rel}(r)$ are the reduced mass and the 
optical potential for the relative motion, respectively. 
The coupling matrix elements, $V_{nm}(r)$, in Eq. (\ref{cceq}) are 
evaluated as follows. 
For the couplings to the collective excitations, we compute the coupling
matrix elements according to the collective model in the full-order coupling
\cite{HT12,ccfull}.
For the couplings to the 
non-collective excitations, on the other hand, 
we employ the random-matrix model~\cite{YHR13}. 
Based on this model, we consider an ensemble of coupling matrix
elements and require that their first moment satisfies
\begin{align}
\overline{V_{nn^\prime}^{II^\prime}(r)} = 0, 
\end{align}
while the second moment satisfies
\begin{align}
&\ \ \ \ \ \overline{V_{nn^\prime}^{II^\prime}(r)V_{n^{\prime\prime}n^{\prime\prime\prime}}^{I^{\prime\prime}I^{\prime\prime\prime}}(r^\prime)} \nonumber \\
&=  \left\{\delta_{nn^{\prime\prime}}\delta_{n^{\prime}n^{\prime\prime\prime}} 
          \delta_{II^{\prime\prime}}\delta_{I^{\prime}I^{\prime\prime\prime}}
   + \delta_{nn^{\prime\prime\prime}}\delta_{n^{\prime}{n^{\prime\prime}}}
                      \delta_{II^{\prime\prime\prime}}\delta_{I^{\prime}{I^{\prime\prime}}}
   \right\} \nonumber
  \\ &\ \ \ \ \times  \sqrt{(2I+1)(2I^\prime+1)}  \nonumber
  \sum_{\lambda}
   \left(
     \begin{array}{ccc}
       I & \lambda & I^{\prime} \\
       0 & 0 & 0
     \end{array}
   \right)^{2} \nonumber\\
&\ \ \ \ \times
\alpha_{\lambda}(n,n^\prime;I,I^\prime;r, r^\prime).
\end{align}
Here, the bars denote an ensemble average. 
$I$ is the spin of the intrinsic state labeled by $n$, 
and $\alpha_{\lambda}$ is the coupling form factor. 

In this paper, for simplicity, 
we take into account the coupling to non-collective
states only from the ground state. This
is similar to the linear coupling approximation.
For the coupling form factor $\alpha_{\lambda}$, 
we assume the following form, 
\begin{eqnarray}
\alpha_{\lambda}(n,0;I,0;r, r^\prime) =
\frac{w_\lambda}{\sqrt{\rho(\epsilon_n)}}
e^{-\frac{\epsilon_n^2}{2\Delta^2}}
e^{-\frac{(r-r^{\prime})^2}{2\sigma^2}}
h(r)h(r^{\prime}),
\label{form_factor}
\end{eqnarray}
where $\rho(\epsilon_n)$ is the level density at excitation energy
$\epsilon_n$, and
($w_{\lambda}, \Delta, \sigma$) are adjustable parameters.
The appearance of the level density in the denominator
of the form factor reflects the complexity of the
non-collective states~\cite{akw1}.
For the function $h(r)$, we assume that it is 
given by the derivative of our Woods-Saxon potential shape, that is,
\begin{eqnarray}
h(r) = \frac{e^{(r-R)/a}}
{\left[1 + e^{(r-R)/a}\right]^2},
\end{eqnarray}
as in the coupling matrix elements for the collective states 
in the linear-coupling approximation. 

\subsection{Potential parameters and the couplings to collective excited states}

For the nuclear potential, we use the Woods-Saxon form with surface
diffuseness parameter $a = 0.65$~fm and radius parameter 
$r_0 = 1.15$~fm for both systems. The depth $V_0$
is taken to be 55.0~MeV for $^{20}$Ne + $^{90}$Zr and 62.3~MeV  
for $^{20}$Ne + $^{92}$Zr. 
The resulting Coulomb barrier heights are $V_{\rm B}$ = 54.0~MeV 
and 53.3~MeV, respectively. 

As for the couplings in the $^{20}$Ne nucleus, 
we consider the rotational 
states in the ground state band up to the 6$^{+}$ state with the deformation
parameters $\beta_2 = 0.46$ and $\beta_4 = 0.27$.
The octupole phonon state 
at 5.62~MeV is also included with $\beta_3 = 0.39$.
For the couplings to the collective excited states in the $^{90}$Zr nucleus,
we take into account the vibrational 2$^{+}$ state at 2.18~MeV with
$\beta_2 = 0.089$ and the 3$^-$ state at 2.75~MeV with 
$\beta_3=0.211$~\cite{KMM10}. For the $^{92}$Zr nucleus, 
we take into account the vibrational 2$^{+}$ state at 0.93~MeV
with $\beta_2$ = 0.103 and the 3$^{-}$
state at 2.34~MeV with $\beta_3 = 0.17$.  
For the quadrupole phonon, following Ref.~\cite{NMD01}, 
we use a slightly larger deformation parameter 
$\beta_2^{(\rm N)}$ = 0.144 for the nuclear coupling. 
These collective excitations in the Zr isotopes are taken into 
account up to the two-phonon states, whereas the mutual excitations 
of the quadrupole and the octupole
phonons are not included.

\subsection{Couplings to non-collective excited states}

\begin{figure}[t]
    \includegraphics[clip,keepaspectratio,width=78mm]{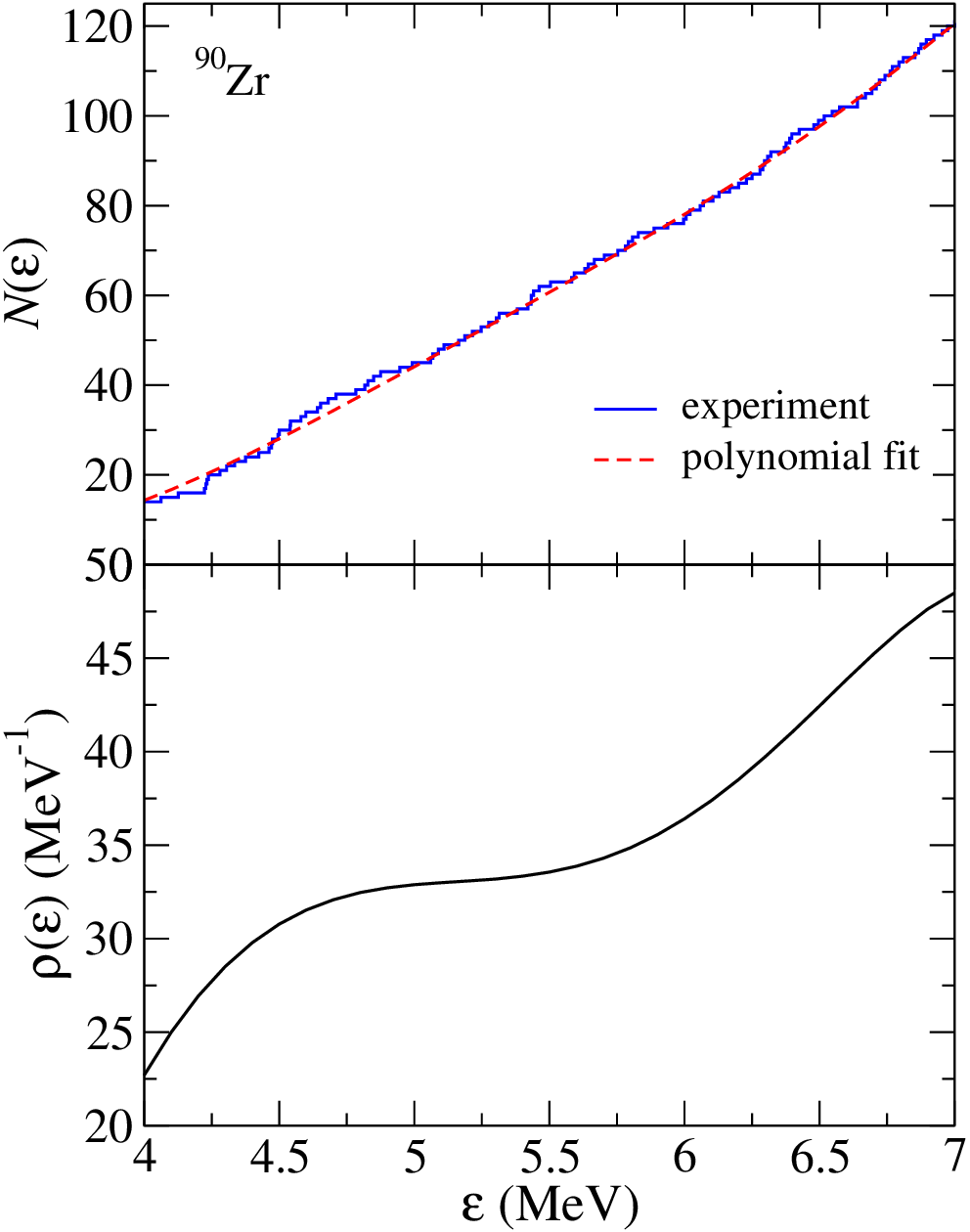}
    \caption{(Color online) Upper panel: the number of levels of 
$^{90}$Zr up to excitation energy
    $\epsilon$ as a function of $\epsilon$. The histogram represents the 
    experimental data~\cite{bnl}, while the dashed line shows its fit with 
a sixth-order polynomial. 
    Lower panel: the continuous level density obtained as the 
first derivative of the
    fitting function of the upper panel.}
    \label{level_fit_90Zr}
\end{figure}

\begin{figure}[t]
    \includegraphics[clip,keepaspectratio,width=78mm]{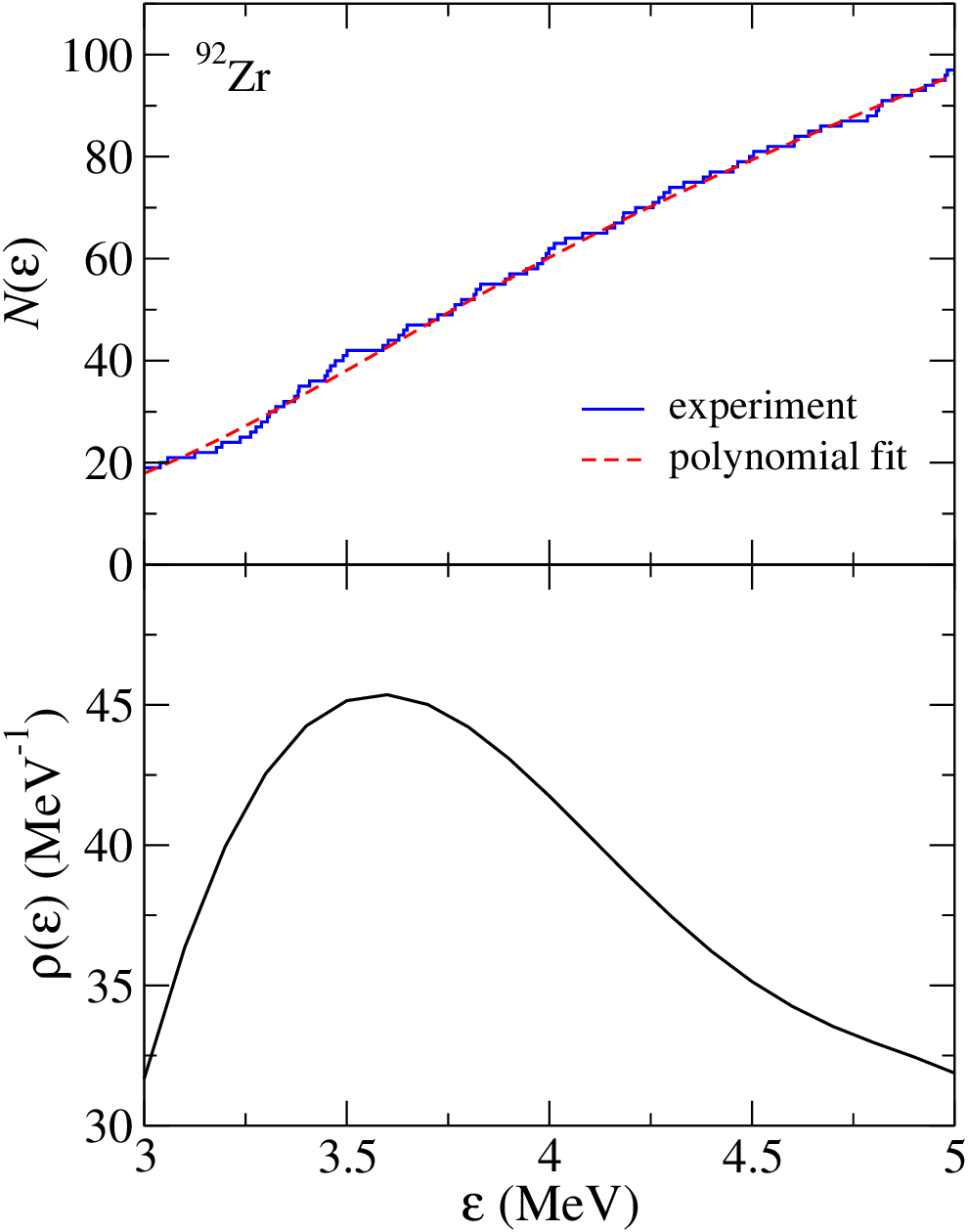}
    \caption{(Color online) The same as Fig. \ref{level_fit_90Zr} but 
for $^{92}$Zr.}
    \label{level_fit_92Zr}
\end{figure}

\begin{table*}[hbt]
\caption{
Coefficients for $n$=0-6 for the 
polynomial fits to the function $N(\epsilon)$ defined by 
Eq.~(\ref{num_levels}). 
The units of $a_n$ are~MeV$^{-n}$. 
}
\begin{center}
\begin{tabular}{c|ccccccc}
\hline
\hline
Nucleus & $a_0$ & $a_1$ & $a_2$ & $a_3$ & $a_4$ & $a_5$ & $a_6$ \\
\hline
$^{90}$Zr & 199.2 & 182.5 & $-$286.5 & 119.7 & $-22$.65 & 2.057 & $-$0.07278 \\
$^{92}$Zr & 63.79 & 540.7 & $-$737.2 & 366.7 & $-$87.00 & 10.07 & $-$04589 \\
\hline
\hline
\end{tabular}
\end{center}
\end{table*}

The aim of this paper is to discuss the effect of noncollecitve 
excitations in the zircronium targets on the 
$^{20}$Ne + $^{90,92}$Zr reactions. We do not consider the noncollective 
excitaions in the $^{20}$Ne projectile, as there exist only a few noncollective 
states in the low-energy region in this light nucleus (for instance, 
the band heads for noncollective bands below 7~MeV are only the 2$^-_1$ state 
at 4.97 MeV and 0$^+_2$ state at 6.73~MeV~\cite{F79,K04}).
For the zirconium non-collective states, the excitation energies and spins 
are well known experimentally 
\cite{bnl}, but the deformation parameters (that is, the coupling strengths) 
are poorly known. We take, therefore, the experimental values of the excitation 
energies and spins, while we estimate the coupling matrix elements  
using the random-matrix model.
Among the non-collective excited states, we take into account only those 
with natural parity, given that both the projectile and the target nuclei 
are even-even nuclei. 
For the parameters in the random-matrix model, we use $\Delta=7$~MeV,
$\sigma=4$~fm, and $w_{\lambda} = w = 200$~MeV$^{3/2}$.
The values for $\Delta$ and $\sigma$ are the same as those in 
Refs.~\cite{akw3,akw4},
while the value for $w$, that determines the coupling strengths to
the non-collective excited states, is chosen 
by fitting the experimental barrier distribution for 
the $^{20}$Ne + $^{92}$Zr system. 
The same values for the parameters 
are then used for the calculations in the $^{20}$Ne + $^{90}$Zr
system, though of course the level density is different 
in that case. 

In order to calculate the coupling matrix elements for the non-collective
excitations, these level densities are required (see Eq. (\ref{form_factor})), 
and in order to implement them in the coupled-channels calculations, 
we introduce a continuous level density as follows~\cite{YHR13}. 
We first note that the level density is defined by
\begin{eqnarray}
\rho(\epsilon) = \sum_n \delta(\epsilon -\epsilon_n),
\end{eqnarray}
for a discrete spectrum.
From the empirical level density $\rho(\epsilon)$, 
we then define the following function, 
\begin{eqnarray}
N(\epsilon) =  \int^\epsilon_0 \rho(\epsilon^\prime)d\epsilon^\prime
=\sum_n \theta(\epsilon -\epsilon_n). 
\label{num_levels}
\end{eqnarray}
This gives the number of levels up to an excitation energy $\epsilon$.
The solid lines in the upper panel of Figs. 1 and 2 show the experimental 
values for $N(\epsilon)$ for $^{90}$Zr and $^{92}$Zr, respectively. 
We next fit this function with a polynomial in $\epsilon$. 
For $^{90}$Zr, we fit $N(\epsilon)$ in the interval 
between 3~MeV and 8~MeV with a sixth-order polynomial
$\displaystyle f(\epsilon) = \sum_{n=0}^{6}a_n \epsilon^n$. 
For $^{92}$Zr, we fit $N(\epsilon)$ in the interval between
2.5~MeV and 6~MeV with a similar function. 
Values of the coefficients $a_n$ are given in Table I. 
The dashed lines in the upper panels of Figs. 1 and 2 show the quality of 
the fits. We then obtain the continuous level densities 
by differentiating $f(\epsilon)$ 
with respect to $\epsilon$. 
The resultant, continuous level densities for the two isotopes 
are shown in the lower panels of Figs. 1 and 2. 

\section{Results}

\subsection{Quasi-elastic scattering cross sections and barrier distributions}

\begin{figure}[t]
    \includegraphics[clip,keepaspectratio,width=78mm]{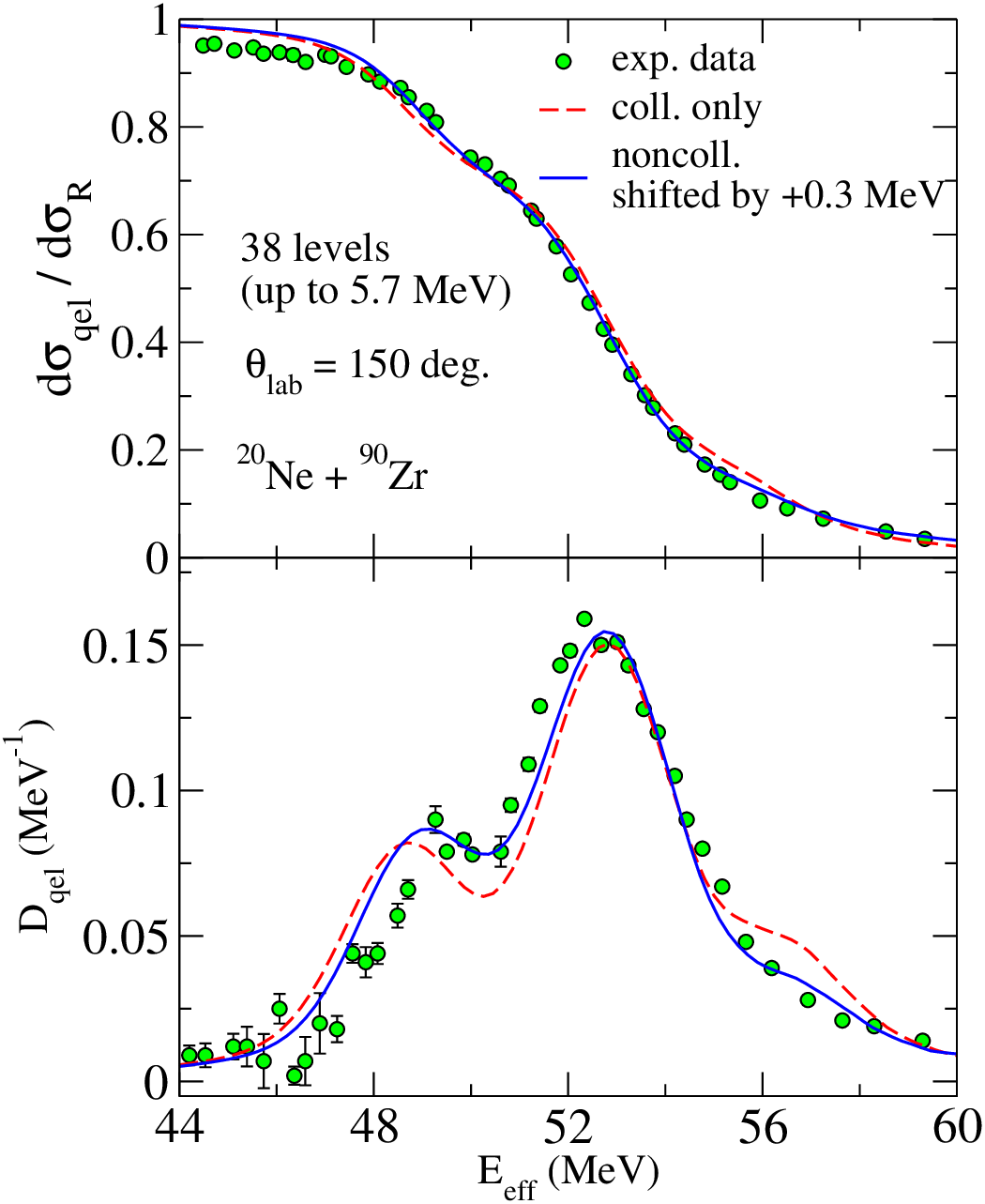}
    \caption{(Color online) The quasi-elastic cross section (upper panel) 
and the quasi-elastic
    barrier distribution (lower panel) for the 
$^{20}$Ne + $^{90}$Zr system at the scattering angle 
$\theta_{\rm lab}=150^{\circ}$.
Dots represent the experimental data, taken from Ref.~\cite{piasecki}. 
Dashed lines show the results obtained by including 
only collective excitations,
while solid lines show the results
including the non-collective excitations. 
The solid lines are shifted in energy by the amount shown in the figure 
in order to compensate for the trivial change of Coulomb barrier height 
due to the non-collective couplings. }
\label{qel_20Ne_90Zr}
\end{figure}

\begin{figure}[t]
    \includegraphics[clip,keepaspectratio,width=78mm]{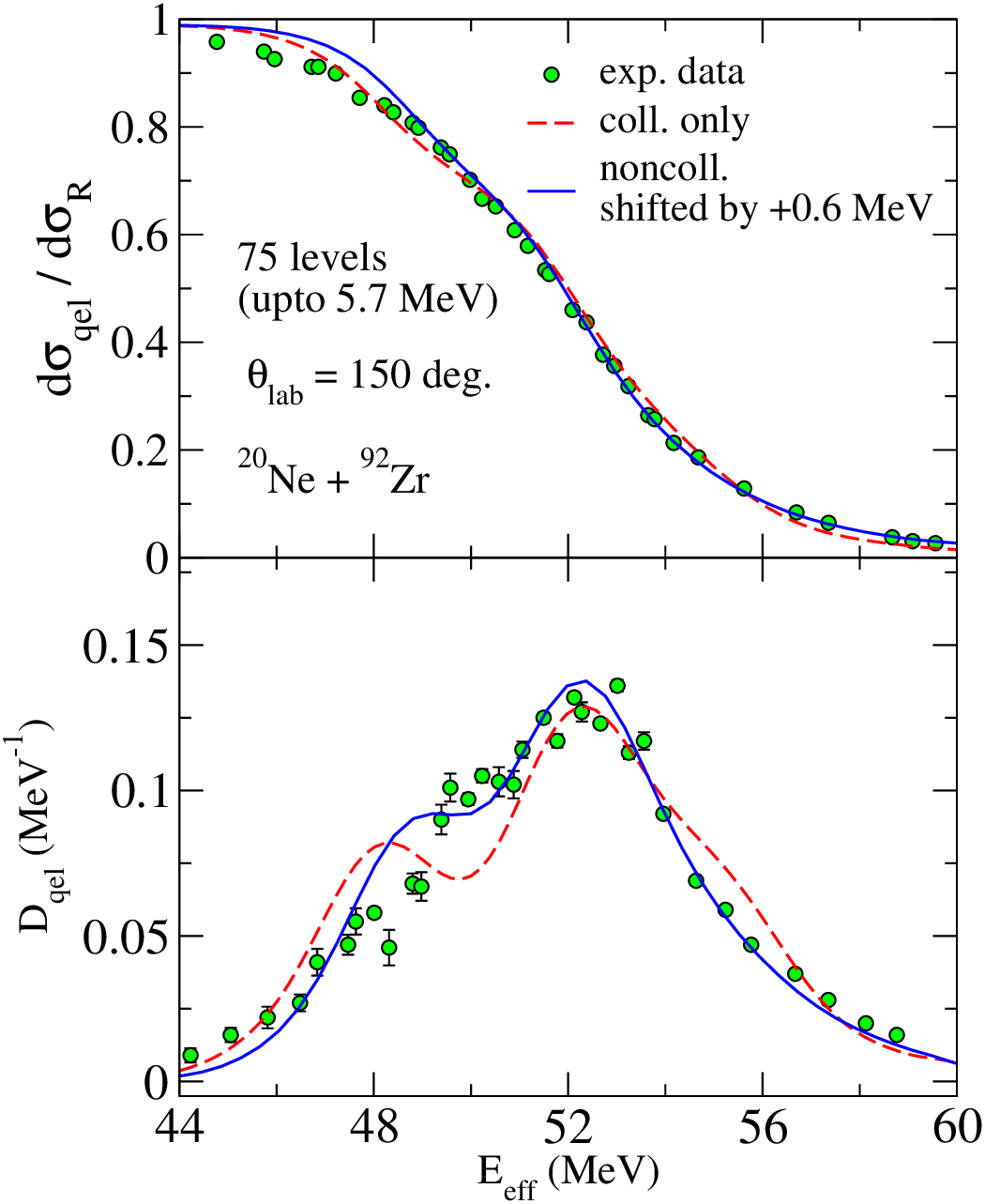}
    \caption{(Color online) The same as Fig. \ref{qel_20Ne_90Zr} 
but for the $^{20}$Ne + $^{92}$Zr system.}
    \label{qel_20Ne_92Zr}
\end{figure}

Let us now solve the coupled-channels equations for the $^{20}$Ne + $^{90,92}$Zr
systems and examine the effect of non-collective excitations on 
their quasi-elastic scattering. 
Fig.~\ref{qel_20Ne_90Zr}
shows the quasi-elastic cross section and 
the quasi-elastic barrier
distribution for the $^{20}$Ne + $^{90}$Zr system,
whereas Fig.~\ref{qel_20Ne_92Zr}  shows the same functions for $^{20}$Ne + $^{92}$Zr.
These quantities are plotted as a function of the effective energy 
defined by
\begin{eqnarray}
E_{\rm eff} = 2E\frac{{\rm sin}(\theta_{\rm c.m.}/2)}
{1+{\rm sin}(\theta_{\rm c.m.}/2)},
\end{eqnarray}
where $\theta_{\rm c.m.}$ is the center-of-mass scattering angle. 
This quantity is introduced to map quasi-elastic cross sections at 
$\theta_{\rm c.m.}$ to their notional values at $\theta_{\rm c.m.}=\pi$, 
by correcting for the centrifugal energy of
the corresponding classical Rutherford trajectory~\cite{timmers,HR04}. 
In both figures, the dots represent experimental data taken
at $\theta_{\rm lab} = 150^\circ$~\cite{piasecki}, while 
the dashed lines show the results that take account of only the collective
excitations. 
The solid lines represent the results that take into account the
non-collective excitations in addition to the collective ones. 
To this end, we include non-collective states up to 5.7~MeV; this 
corresponds to 38 levels in $^{90}$Zr and 75 levels in $^{92}$Zr. 
We have confirmed that the results do not change significantly if 
the non-collective states are truncated at 
higher excitation energies. Note that 
these results have been obtained with a single realization of the coupling 
matrix elements. In principle one should repeat the calculations many 
times with randomly generated matrix elements and
take an ensemble average.  However, we have verified in a smaller model space 
that the dispersion due to the randomness of these
elements is sufficiently small that
a single realization already yields reasonable results. 
Note also that the excitation energies of
the non-collective states are relatively large, so 
that coupling to them leads to an adiabatic renormalization  
of the barrier~\cite{HT12,THAB94}.
We have therefore shifted the solid lines in energy by +0.3 and +0.6~MeV 
for the $^{20}$Ne + $^{90}$Zr and $^{20}$Ne + $^{92}$Zr systems, respectively, 
in order to compensate for this trivial modification of the barrier height. 

For the $^{20}$Ne + $^{90}$Zr reaction, one sees that 
non-collective excitations do not alter the barrier distribution
in a significant way, though the
dip between the two main peaks is somewhat filled.
In marked contrast, in the $^{20}$Ne + $^{92}$Zr reaction, 
the non-collective excitations almost completely fill this dip, 
leading to a much more smeared barrier distribution. 
Furthermore the overall width of the distribution
becomes smaller in this case. 
These changes all lead to a much more satisfactory 
agreement with the experimental data. 
Moreover, we note that the non-collective excitations considerably 
improve the behavior around $E_{\rm eff}\sim 56$~MeV 
in both systems. 
In all these calculations, the same values for the parameters 
$w$, $\Delta$, and $\sigma$ of the random-matrix model are used. 
Therefore, any difference in the non-collective  effects 
originates solely from 
the different level densities of these zirconium isotopes.
That is, the effect of non-collective excitations is greater for 
$^{92}$Zr since a larger number of non-collective states 
exist in the region of relatively-low excitation energy. 
As noted above, this higher level density is
due to the two extra neutrons outside
the $N = 50$ closed neutron shell in $^{90}$Zr. 

Our calculations also indicate 
that the role of noncollective excitations in {\it fusion} 
barrier distributions is similar to that in the {quasi-elastic} 
barrier distributions. That is, the fusion barrier distribution 
for the $^{20}$Ne+$^{92}$Zr system is significantly altered due to 
the noncollective excitations, 
in a similar manner as in the corresponding quasi-elastic 
barrier distribution shown in Fig. 4, 
while the fusion barrier distribution for 
the $^{20}$Ne+$^{90}$Zr system is not affected much. 
This, of course, is rather expected since there is no reason why 
the different couplings for the two Zr isotopes should not affect 
the fusion and quasi-elastic barrier distributions in a different 
way. This suggests that one can indeed have a smoother fusion 
barrier distribution for the $^{20}$Ne+$^{92}$Zr system 
as compared to that for the $^{20}$Ne+$^{90}$Zr system. 

The calculations shown in Figs. 3 and 4 
still do not reproduce the 
quasi-elastic scattering cross sections at low energies 
around $E_{\rm eff}\sim$ 46~MeV, despite the fact that the calculations 
agree well with the data at higher energies. 
We do not know the origin of this discrepancy, 
but other effects, such as $\alpha$ pick-up reactions, 
might play some role. 

\subsection{$Q$-value distribution}

\begin{figure}[t]
  \center
    \includegraphics[clip,keepaspectratio,width=80mm]{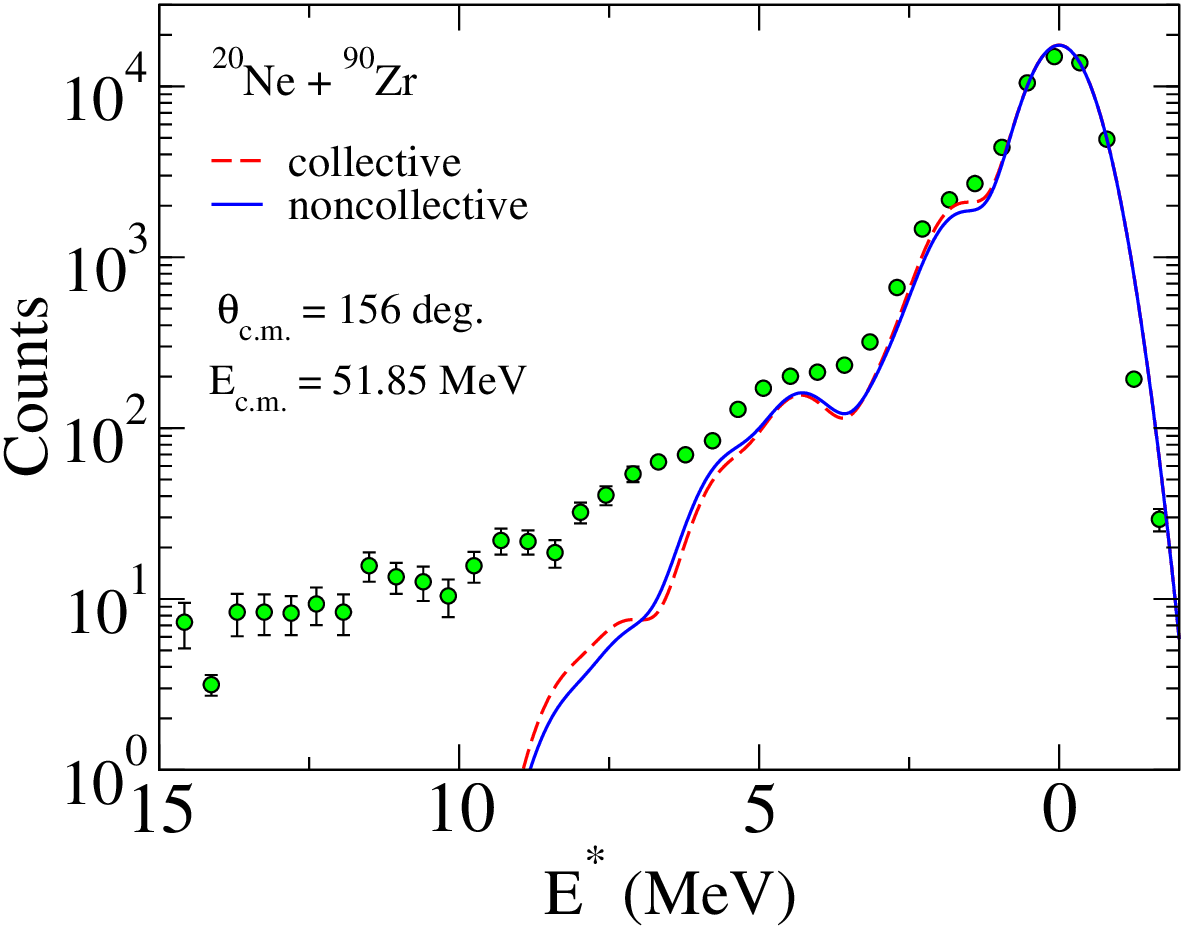}
    \caption{(Color online) The $Q$-value distributions for 
the $^{20}$Ne + $^{90}$Zr system at a scattering angle  
$\theta_{\rm c.m.}=156^{\circ}$ and incident center-of-mass energy 
$E_{\rm c.m.}=51.85$~MeV. 
Experimental data are taken from Ref.~\cite{piasecki}.
The calculated results are smeared with a Gaussian function with 
a width of $\eta$=0.5~MeV, and then normalized to the experimental 
elastic peak at $E^*$ = 0. }
\label{fig7.15}
\end{figure}

\begin{figure}[t]
  \center
    \includegraphics[clip,keepaspectratio,width=80mm]{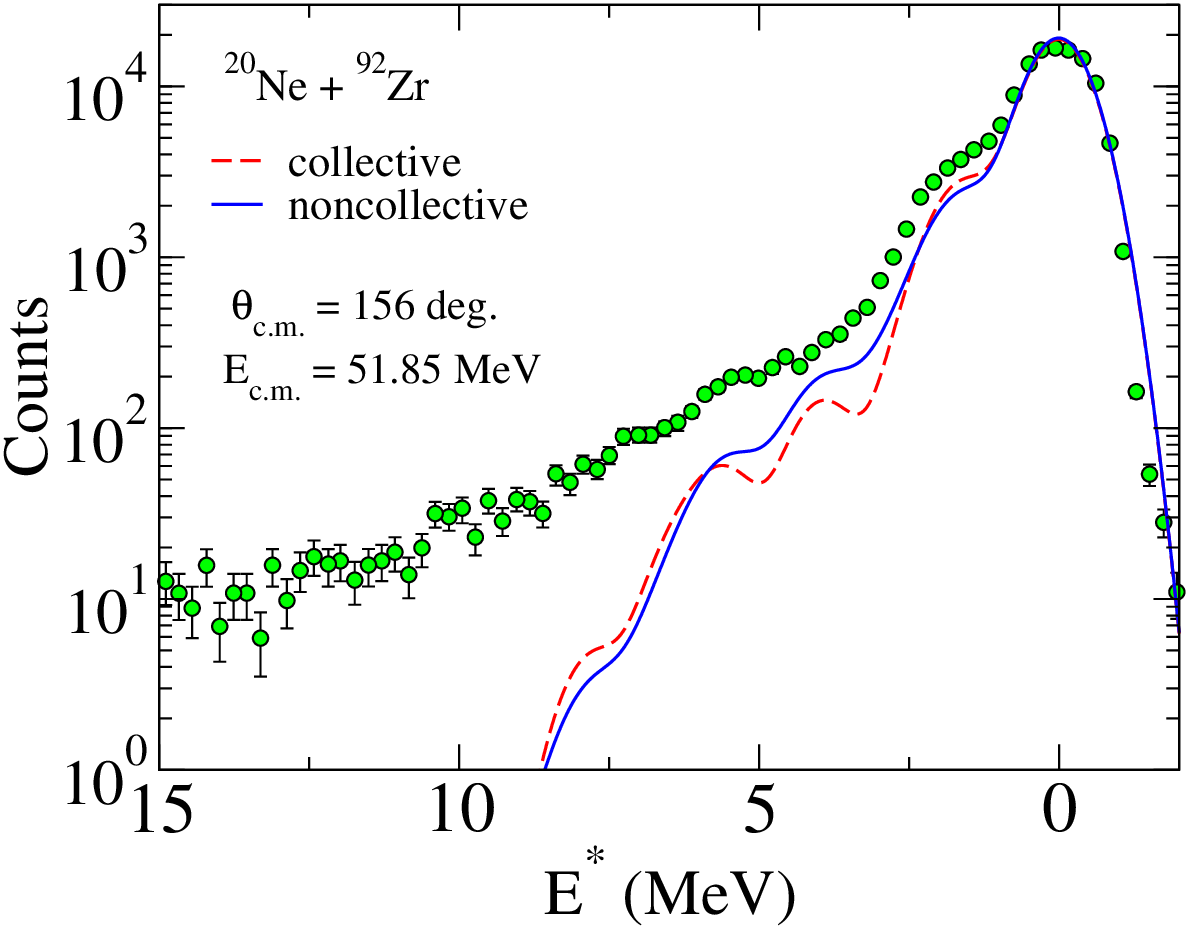}
    \caption{(Color online) The same as Fig. 5, but for 
the $^{20}$Ne + $^{92}$Zr system. 
    Experimental data are from Refs.~\cite{piasecki,piasecki-priv}.}
    \label{fig7.16}
\end{figure}

\begin{figure}[t]
  \center
    \includegraphics[clip,keepaspectratio,width=80mm]{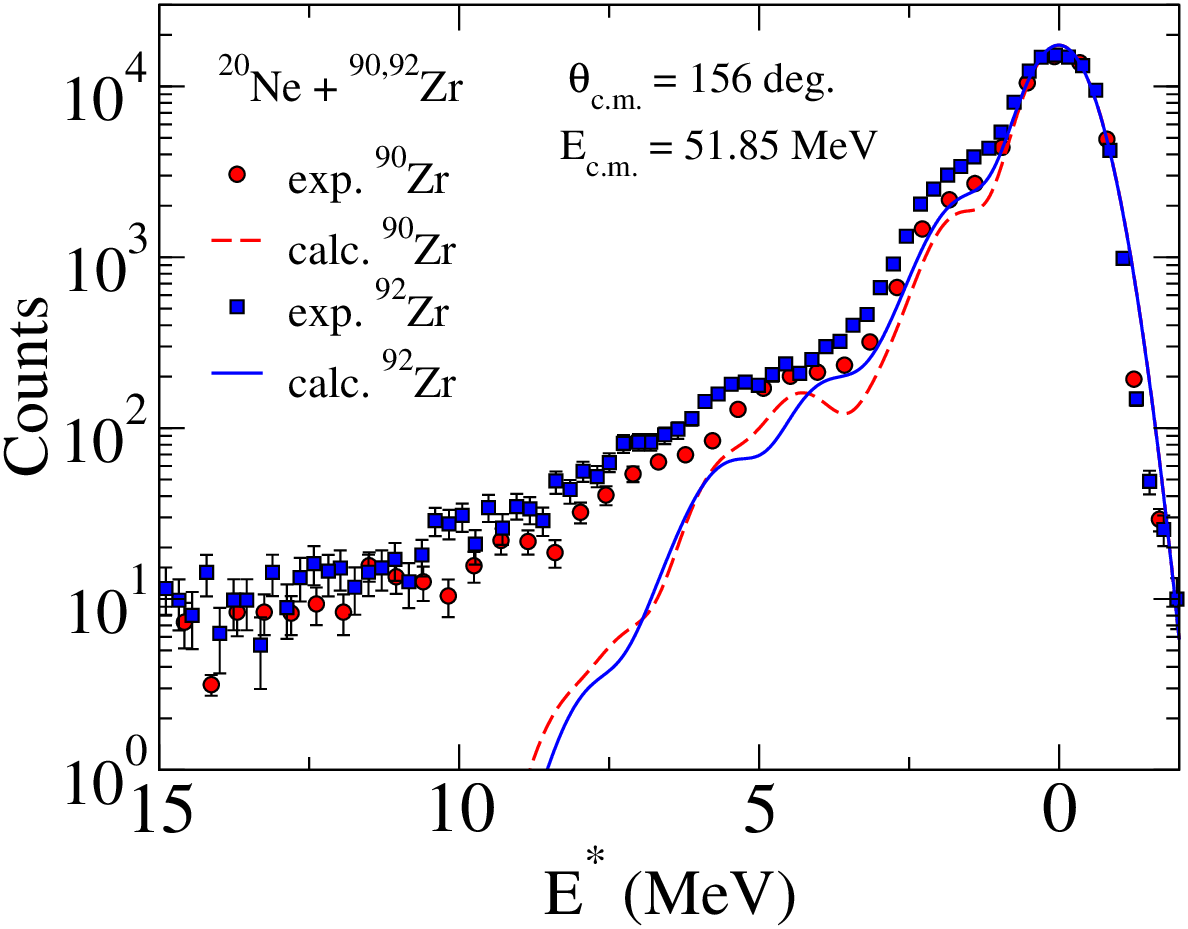}
    \caption{(Color online) Comparison of the $Q$-value distributions 
for the $^{20}$Ne + $^{90}$Zr
and $^{20}$Ne + $^{92}$Zr systems. The circles and the dashed line show, 
respectively,the experimental data and the calculated result for the 
$^{20}$Ne + $^{90}$Zr system, while the squares and the solid 
line represent the $^{20}$Ne + $^{92}$Zr system.}
\label{qdist_comp}
\end{figure}

We next discuss the $Q$-value distribution, that is, the 
excitation energy spectra. Figures \ref{fig7.15} and \ref{fig7.16} 
show the $Q$-value distributions 
for the $^{20}$Ne + $^{90}$Zr and $^{20}$Ne + $^{92}$Zr systems,
respectively.
The meaning of each line is the same as in Figs. 3 and 4. 
The experimental data were taken at $\theta_{\rm c.m.} = 156^{\circ}$ and
$E_{\rm c.m.} =51.85$~MeV and do not include the transfer 
cross sections~\cite{piasecki,piasecki-priv}.
The theoretical $Q$-value distributions are evaluated at 
$E$=51.55 and 51.25 MeV 
for the $^{20}$Ne + $^{90}$Zr and $^{20}$Ne + $^{92}$Zr systems,
respectively, that is, at those energies 
corresponding to 
$E_{\rm c.m.} =51.85$~MeV after the energy shifts indicated in Figs. 3 and 4 
are taken into account. They are 
obtained by summing over the different channels
as follows 
\begin{eqnarray}
F(E^{*}) \propto 
\sum_{n}\frac{d\sigma_n}{d\Omega}\,\frac{1}{\sqrt{2\pi}\eta}
e^{-\frac{(E^*-\epsilon_n)^2}{2\eta^2}}, 
\label{eq5.6}
\end{eqnarray}
that is, we smear with a Gaussian function of width $\eta$
to simulate the experimental energy resolution. 
The normalization factor and the value of the width 
($\eta = 0.5$~MeV) are determined so that the elastic peak 
in the experimental $Q$-value distribution is reproduced. 

One can see that 
the non-collective excitations
affect little the $Q$-value distribution at this incident energy 
for either system. 
For $^{20}$Ne + $^{90}$Zr, the calculation reasonably
reproduces the data up to about $E^*= 5$~MeV, 
although it underestimates the experimental data at higher energies due to the 
truncation of the non-collective states in our calculations. 
For the $^{20}$Ne + $^{92}$Zr system, 
the non-collective excitations somewhat enhance the contribution 
from the inelastic
channels between about 3 to 6~MeV, and the experimental data are 
reasonably well reproduced up to 4~MeV.

Figure \ref{qdist_comp} compares the $Q$-value distributions for the two 
systems. The circles and the dashed lines are for 
the $^{20}$Ne + $^{90}$Zr system, while the squares and the solid lines 
are for the $^{20}$Ne + $^{90}$Zr system.
They are all normalized to the height of the elastic peak in the 
experimental data for the $^{20}$Ne + $^{90}$Zr system.
One can see that the experimental $Q$-value distributions are 
similar for both the systems, and are well reflected 
by the present coupled-channels calculations. 

\begin{figure}[t]
  \center
    \includegraphics[clip,keepaspectratio,width=85mm]{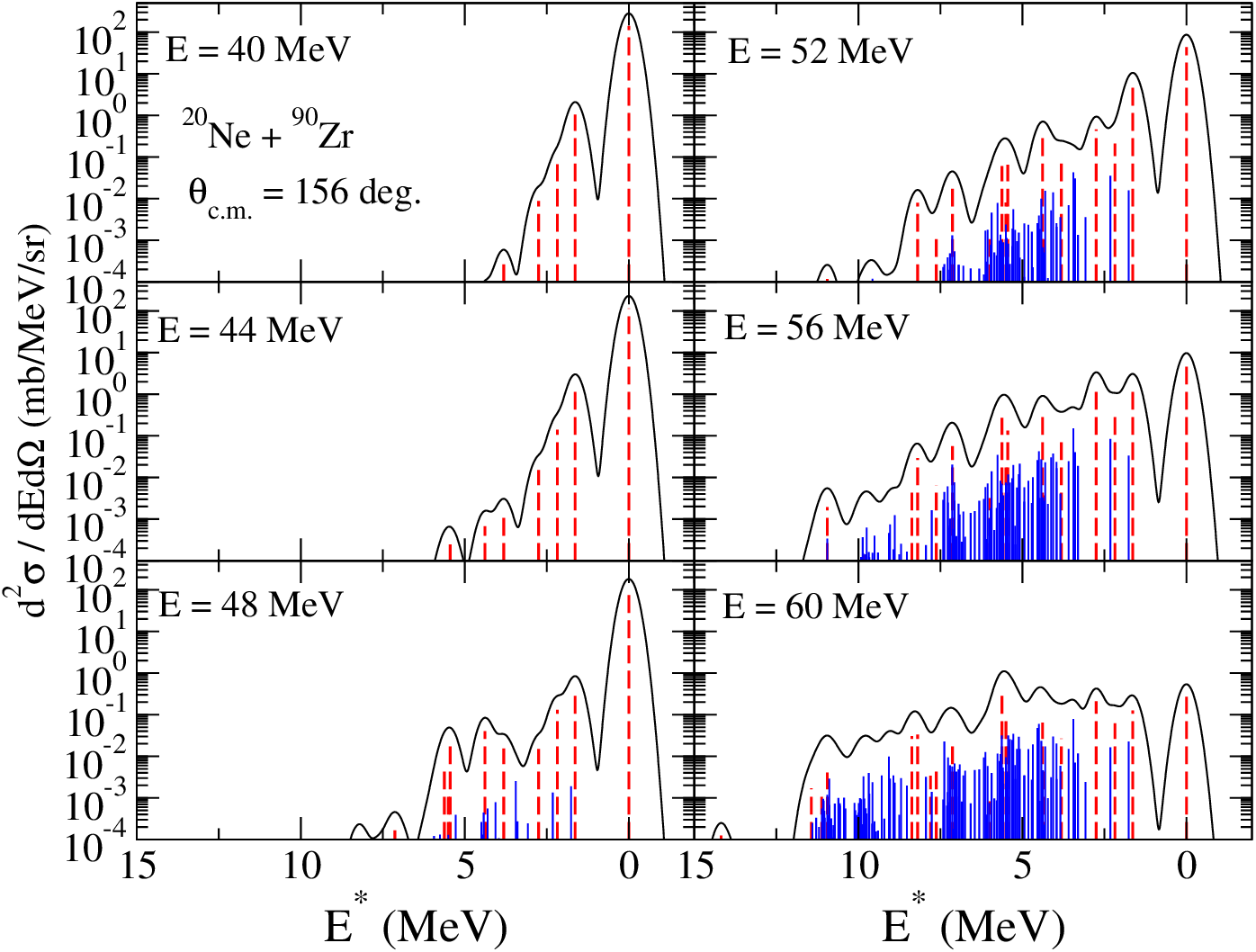}
    \caption{(Color online) The energy dependence of the $Q$-value 
distribution for the $^{20}$Ne +
$^{90}$Zr system. The dashed and the
solid peaks show the contributions from the collective and the
non-collective excitations, respectively. The solid lines are obtained by
smearing the spectra with a Gaussian function of width 0.2~MeV.}
\label{fig7.21}
\end{figure}

\begin{figure}[t]
  \center
    \includegraphics[clip,keepaspectratio,width=85mm]{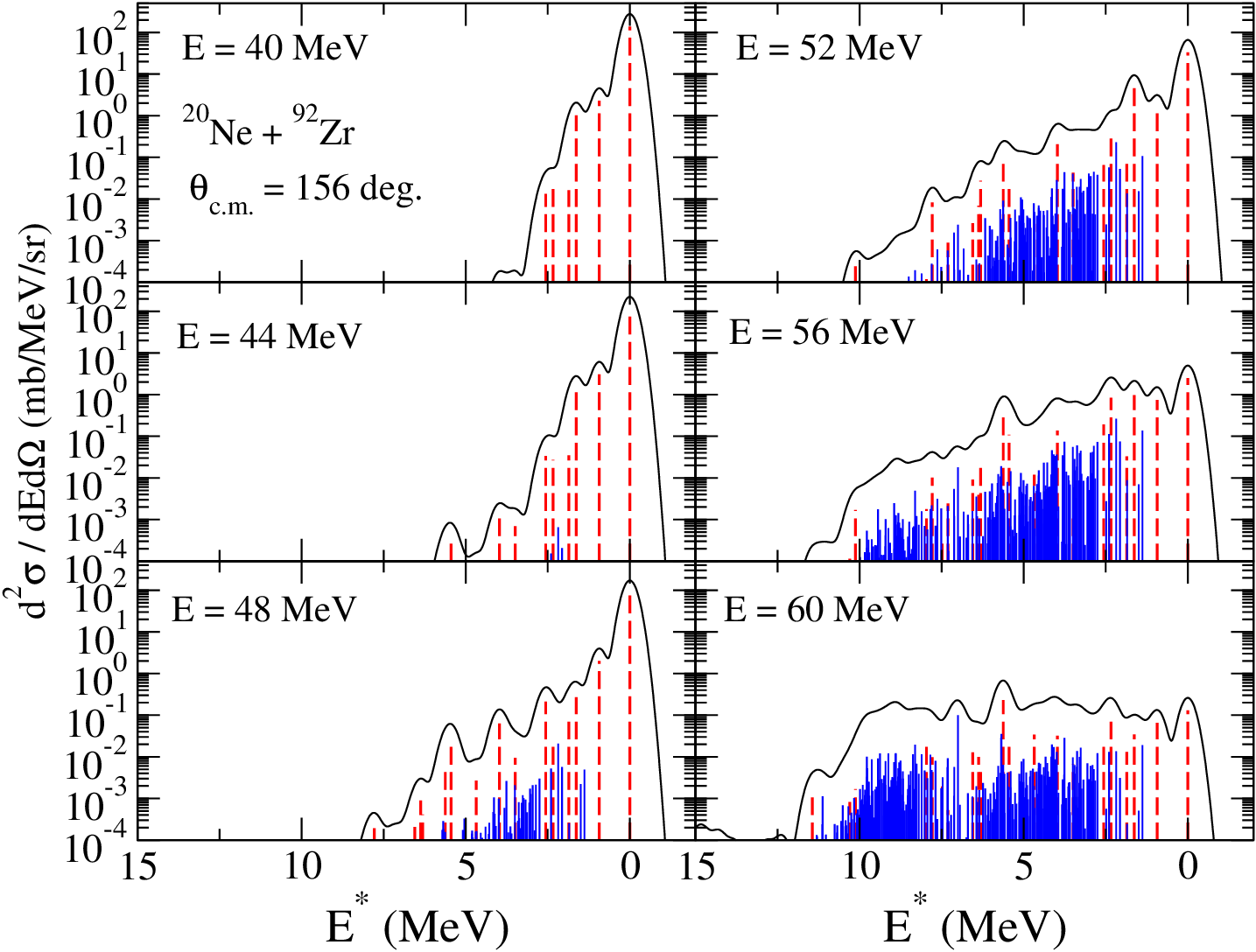}
    \caption{(Color online) Same as Fig.\ref{fig7.21}, but for 
$^{20}$Ne + $^{92}$Zr.}
    \label{fig7.22}
\end{figure}

This might appear surprising given that there is a large difference
between the two measured quasi-elastic barrier distributions 
(see Fig. 3 and 4). This does not necessarily 
mean that the non-collective excitations do not play an important role 
in the $Q$-value distribution, however. 
In order to demonstrate this, 
we show in 
Figs. \ref{fig7.21} and \ref{fig7.22} the energy dependence of the
$Q$-value distribution 
obtained at 
different incident energies from 40~MeV to 60~MeV 
for the $^{20}$Ne + $^{90}$Zr and $^{20}$Ne + $^{92}$Zr
systems.
The dashed peaks show contributions from the
collective channels, 
while the solid peaks show the contributions from the
non-collective channels. 
We also show envelopes of the peaks by the solid lines, that 
are obtained by smearing with a Gaussian function 
of width 0.2~MeV.
For both systems, the contribution from the elastic channel and the 
collective excitations is dominant at energies below the barrier, 
while the contribution from the non-collective
excitations becomes more important as the incident energy increases.
This tendency was also observed in our previous calculations for
the $^{16}$O + $^{208}$Pb system~\cite{YHR12} (see 
Refs.~\cite{evers,lin,evers2} for the corresponding experimental data). 
We note again that in the present systems, the non-collective 
excitations contribute more in the $^{20}$Ne + $^{92}$Zr system, and 
it would therefore be interesting to compare the experimental
$Q$-value spectra for the two systems at higher energies than 
studied in Ref.~\cite{piasecki}. There the effect of non-collective 
excitations might be seen more clearly. 

\section{Summary}

We have investigated the role of non-collective excitations of
Zr isotopes in the $^{20}$Ne + $^{90,92}$Zr reactions. This was 
motivated by
recent quasi-elastic scattering experiments for these systems, in which 
the conventional coupled-channels
calculations could not explain the difference between the two 
quasi-elastic barrier distributions. 
In this paper, we have employed the random-matrix model to generate 
appropriate couplings to non-collective states, enabling us to include these
excitations in our coupled-channels calculations.

The results indicate that these
excitations fill in the dip between the
two main peaks in the barrier distribution for the 
$^{20}$Ne + $^{92}$Zr system, considerably smearing its peak structure. 
In contrast, the effect is much smaller for $^{20}$Ne + $^{90}$Zr, 
and the peak structure is not greatly affected by the inclusion of
the non-collective excitations.
The difference
arises solely from the different level densities in these Zr isotopes.
That is, the number of low-lying, non-collective
states is much larger in $^{92}$Zr than in $^{90}$Zr. 
In both systems, the agreement with the experimental 
data for the quasi-elastic scattering
cross sections and the barrier distribution is improved 
by the inclusion of these excitations. 

We have also calculated the $Q$-value distribution for 
$^{20}$Ne + $^{90,92}$Zr scattering. At an incident energy  
$E_{\rm c.m.}$ = 51.85~MeV, where experimental data exist, 
our calculations indicate that 
the contribution from the non-collective excitations
is relatively small, even in the $^{20}$Ne + $^{92}$Zr system. 
In fact, the data show that the $Q$-value distributions do 
not differ significantly at this energy, a result consistent 
with our calculations. 
We have also calculated the energy dependence of the $Q$-value 
distribution
for both systems, and have found that the contribution
from the non-collective excitations becomes more important
as the incident energy increases. 
A similar tendency has been observed experimentally in 
the $^{16}$O + $^{208}$Pb system. 

Non-collecive excitations are expected to become more important 
as the mass number increases (notice that the 
effect of noncollective excitations appear to be larger in the 
$^{16}$O+$^{208}$Pb reaction ~\cite{YHR12} than in the 
$^{20}$Ne + $^{90}$Zr reaction despite the fact that both $^{208}$Pb and $^{90}$Zr 
are closed-shell nuclei. This is party because the effect is somewhat 
amplified in the former due to the larger charge product.) 
In this respect, we 
mention that the random-matrix model employed in this paper 
may be useful for the study of heavy-ion, deep-inelastic collisions, 
where a large number of non-collective excitations play a role.
Even though the random-matrix model has been applied to 
deep-inelastic collisions in the 1970's 
by Weidenm\"uller {\it et al.}, 
a major difference in our work is that we have solved the 
coupled-channels equations quantum mechanically;
this is essential for low-energy heavy-ion reactions.
An interesting future problem would be to develop a more quantum 
mechanical approach for the deep-inelastic processes, still based 
on the random-matrix model. 

\begin{acknowledgments}
We thank E. Piasecki for useful discussions. 
This work was supported by the Global COE Program
``Weaving Science Web beyond Particle-Matter Hierarchy'' at
Tohoku University, and by the Japanese
Ministry of Education, Culture, Sports, Science and Technology
by Grant-in-Aid for Scientific Research under
the program number (C) 22540262.
\end{acknowledgments}

\end{document}